\documentclass[aps,onecolumn,prd,showpacs,showkeys,preprintnumbers,superscriptaddress,nobibnotes,notitlepage,floatfix,longbibliography,nofootinbib]{revtex4-2}

\pdfoutput=1

\usepackage{graphicx}
\usepackage{natbib}
\usepackage{hyperref}
\usepackage{cleveref}
\usepackage{xcolor}
\usepackage{gensymb}
\def\gev{{\rm GeV}}

\def\pev{{\rm PeV}}

\begin{document}

\title{Are There Correlations in the HAWC and IceCube High Energy Skymaps Outside the Galactic Plane?}

\author{Jason Kumar}
\affiliation{Department of Physics and Astronomy, University of Hawai'i, Honolulu, HI 96822, USA}

\author{Carsten Rott}
\affiliation{Department of Physics and Astronomy, University of Utah, Salt Lake City, UT  84112, USA}
\affiliation{Department of Physics, SungKyunKwan University, Jang-an-gu, Suwon 16419, Korea}

\author{Pearl Sandick}
\affiliation{Department of Physics and Astronomy, University of Utah, Salt Lake City, UT  84112, USA}

\author{Natalia Tapia-Arellano}
\affiliation{Department of Physics and Astronomy, University of Utah, Salt Lake City, UT  84112, USA}

\begin{abstract}
We use publicly-available data to perform a search for correlations of high energy neutrino candidate events detected by 
IceCube and high-energy photons seen by the HAWC collaboration. Our search is focused on unveiling such correlations outside of the Galactic plane. This search is sensitive to correlations 
in the neutrino candidate and photon skymaps which would arise from a population of unidentified point sources.  We find 
no evidence for such a correlation, but suggest strategies for improvements with new data sets.
\end{abstract}

\maketitle

{\it Introduction.}
A variety of observatories are currently extending our understanding of astrophysics 
at high energy.  The IceCube Neutrino Observatory has identified thousands of astrophysical neutrino candidate events, each of which could be a first glimpse of a future identification of a neutrino point source~\cite{IceCube:2021xar}. To identify such sources, searches have been performed for spatial and/or temporal clustering of events, in addition to multi-messenger approaches which attempt to correlate high-energy gamma rays  with neutrino events  \cite{Queiroz_2016,IceCube:2022osb,Negro:2023kwv,ANTARES:2020zng}.  Latter 
studies have been motivated by the fact that neutrinos and photons both point back to their astrophysical sources, 
as they are not bent by galactic magnetic fields. 
Moreover, 
weak processes which can produce high-energy neutrinos (such as the decays 
of highly boosted charged pions)
are often related to electromagnetic processes which can produce high-energy gamma rays (such as the decays of highly boosted 
neutral pions)~\cite{Queiroz:2016zwd}, 
implying that a 
source of high-energy neutrinos may also be bright in high-energy gamma rays.  
In this work, we determine if a correlation can be found between unidentified neutrino point sources 
and unidentified gamma-ray sources, using publicly-available data from IceCube and HAWC.
Although attenuation due to pair-production against extragalactic background light (EBL) and the Cosmic Microwave Background (CMB) implies that the sources of higher energy gamma-rays observed by HAWC are likely galactic, lower energy photons observed by HAWC could certainly originate with extragalactic sources~\cite{Gilmore_2009,Vernetto:2016alq}.  As a 
result, our search is sensitive to both extragalactic and Galactic unidentified point source 
populations that are outside of the Galactic plane.

We focus on a sample of 139 neutrino candidate events with energies $\geq \pev$ which lie outside of the Galactic plane 
and within the HAWC footprint, identified by IceCube during the time 
period of May 13, 2011 to July 08, 2018. The publicly-available data released by IceCube includes neutrino candidate events in the range $( \sim 10^3, 10^7)$ GeV, observed between 2008 and 2018.  This sample was released as a result of the analysis of the blazar TXS 0506+056 with the intention to perform multi-messenger searches similar to the one presented here. We restrict our analysis to the last 7 years, after detector completion, and to energies above 1 PeV. The subtraction of the Galactic plane 
($|b| <20\degree$)
results in the 139 events considered in this analysis. Using the publicly-available HAWC point source search sky map employed in the creation of the HAWC catalog \cite{HAWC:2020hrt}, we compare the 
combined test statistic for point sources in these regions with the distribution of combined test statistics 
for a sample of similarly chosen random regions.   We find that the combined point source test statistic for 
the regions which contain high energy neutrino candidates lies well within the 68\% containment region, indicating no 
significant correlation between high-energy neutrino candidates detected by IceCube and gamma rays detected by HAWC.

The plan of this letter is as follows.  We begin by discussing the motivation for this work.  We then describe 
the methodology and results of the analysis.  Finally, we conclude.

{\it Motivation.}  
There are good reasons to suspect that sources of high-energy neutrinos may also be sources of 
high-energy gamma rays.  Examples of sources which have been found to emit both high-energy neutrinos and 
gamma-rays include the blazar TXS 0506+056~\cite{IceCube:2018dnn,IceCube:2018cha} and the 
Seyfert galaxy NGC 1068~\cite{IceCube:2022der}.  

As a general rule, processes which can produce neutrinos are often related to processes which 
can produce photons.  As an example, neutrinos can be produced by particle physics processes which produce 
highly boosted charged pions ($\pi^\pm$), which decay via $\pi^\pm \rightarrow \mu^\pm \nu$.  But since isospin is 
an approximate symmetry, processes that produce charged pions are typically accompanied by processes producing 
neutral pions ($\pi^0$), which decay to photons via $\pi^0 \rightarrow \gamma \gamma$. 

More generally, though, any hard processes which produce high-energy neutrinos are necessarily accompanied by processes in 
which weak gauge bosons ($W^\pm,Z$) are emitted from the neutrino via bremsstrahlung.  These weak gauge bosons will decay 
with an ${\cal O}(1)$ branching fraction to hadrons, with hadronic decays also producing photons.  These weak 
bremsstrahlung processes are suppressed by factors of the weak coupling constant, but  because photons are much 
easier to detect than neutrinos, the photon production processes may 
be competitive with the high-energy neutrino production process that they accompany~\cite{Queiroz_2016}.  
We thus see that, even if 
high energy neutrinos are produced by beyond-the-Standard-Model (BSM) physics, such as dark matter annihilation, 
we should generally expect an associated photon signal.

{We have described a few processes in which both photons and neutrinos are produced, with related spectra. However, photons and neutrinos may be produced through multiple processes that are not directly related to each other. As one example, dark matter particles may annihilate into multiple Standard Model final states. The subsequent decays of these Standard Model particles produce photons and neutrinos, along with other stable particles.  The relationship 
between the photon energy spectrum and the neutrino energy spectrum would then depend in detail not only on the dark 
matter mass, but also on the branching fractions to the various final states.  As another example, a source of high-energy 
neutrinos may also produce other cosmic rays, such as high-energy electrons or positrons, which can produce gamma rays 
through inverse Compton scattering off background photons.  }

{Moreover, even the rough energy scale of the gamma rays at production, in comparison to the neutrino energy 
scale, depends on the production mechanism.  For example, if photons and neutrinos are produced by the decays of neutral 
and charged pions, respectively, then their energy scales will tend to be comparable.  On the other hand, if a 
process such as dark matter annihilation produces neutrinos, then the photons produced as a result of electroweak bremsstrahlung 
will be suppressed by in energy by up to an order of magnitude~\cite{Queiroz_2016}.}

{Furthermore, the photon spectrum will generally be reprocessed by interactions, both at the source and along 
the line of sight to the detector.  For example, the high-energy photons can generate electromagnetic cascades by scattering  
off intervening background light (including starlight, CMB photons, etc) to produce $e^+ e^-$ pairs, which in turn produce 
gamma-rays by inverse Compton scattering off background light.  As such, the photon spectrum may be degraded in energy
in comparison to that of the neutrinos.  But how much reprocessing occurs will also depend on distance from the source to the 
Earth, as well as the details of the environment of the source.}

{We thus see that, given a specific model for the processes which produce neutrinos and photons at 
the source, one can obtain a prediction for the gamma ray spectrum, which can be searched for in data.  But if one is 
agnostic about the production mechanism, one more generally sees the possibility that objects which emit high-energy 
neutrinos are also sources of gamma rays, as well as other cosmic rays.  But as neutrinos and gamma rays are the messengers 
which will both 
point back to the source, there may be a correlation between the photon and neutrino skymaps.  It is this possibility which 
we will study.}

Previous studies have searched for correlations between a neutrino flux and a gamma ray flux when 
one or both of these messengers are produced by identified point sources~\cite{IceCube:2019cia,IceCube:2022zbd}.  
However, there are likely many neutrino point sources which have remained hidden. Due to the inherent difficulty of detecting neutrinos, the number of 
neutrinos detected from each of these sources is small.  
We only consider the most energetic neutrino candidate events, which carry a higher probability of being neutrinos of astrophysical origin, and utilize events that are reconstructed as track-like, the signature of muon neutrinos, for good pointing accuracy. For our analysis we consider these muon neutrino candidates from the Southern hemisphere, observed as ``down-going'' events by IceCube, with reconstructed energy $E \geq \pev $.
Assuming each high-energy neutrino candidate is a potential source,
we determine if there is a statistical excess 
of high energy gamma rays observed by HAWC~\cite{historical:2023opo} from these regions, compared to a similar sample of random sky regions.  This analysis could 
allow us to detect a correlation between neutrinos and gamma rays which, though insignificant in any single 
sky region, could be statistically significant when summed over all such regions.  A similar search for correlations between 
the unresolved gamma-ray background and neutrino events was considered in the context of Fermi-LAT gamma-ray data in the 
$1-25~\gev$ range~\cite{Negro:2023kwv}.  Our analysis will be sensitive to correlations at higher energies.  
Although there are reasons to expect that high energy astrophysical neutrinos are accompanied by much 
lower energy photons (see~\cite{Murase:2015xka, Fang:2022trf}), it is important{, for the reasons discussed above,}
to test the possibility that 
higher energy photons from these sources also arrive at Earth.

\begin{figure}[h!]
    \centering
    \includegraphics[width=0.8\textwidth]{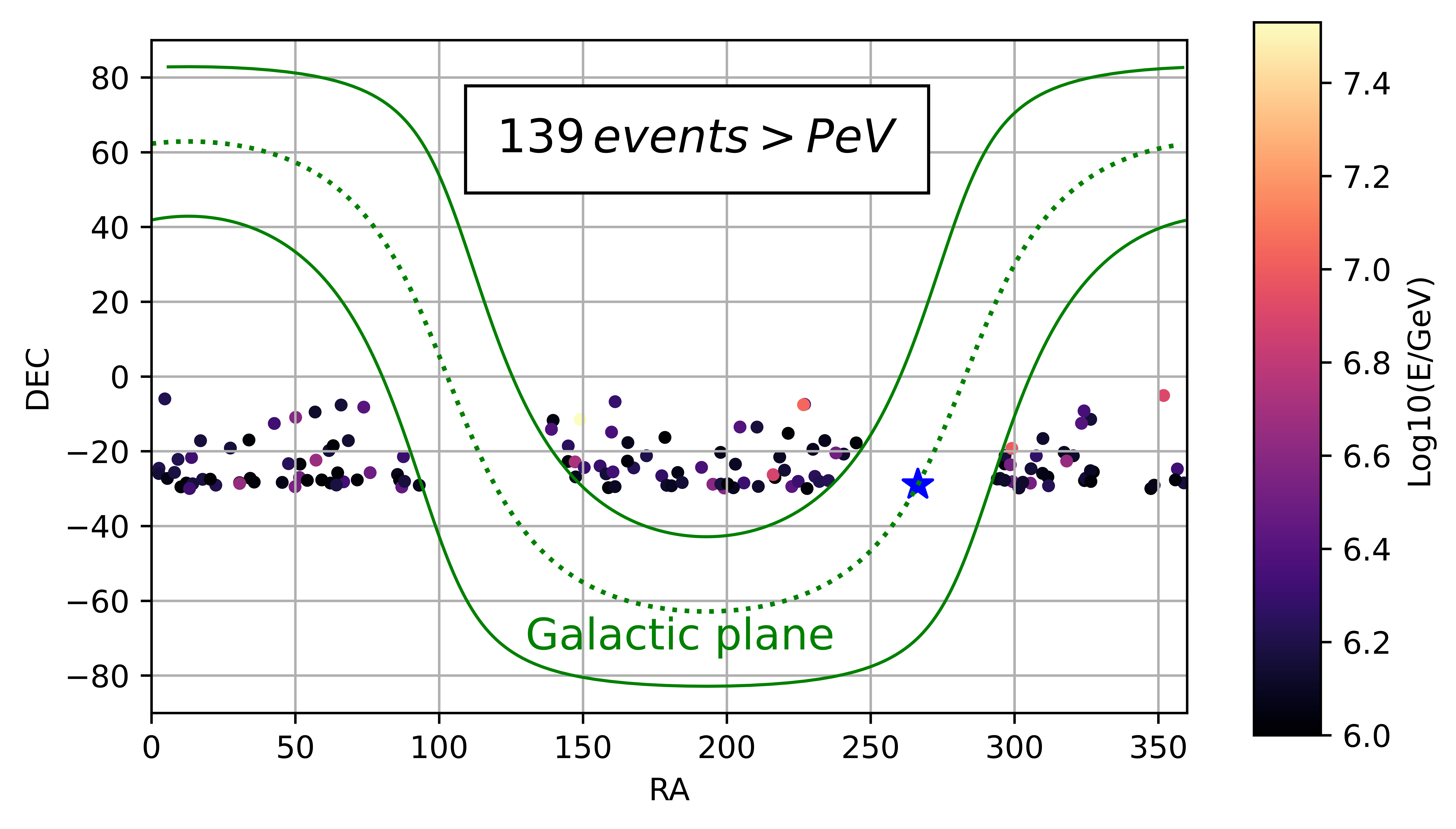}
    \caption{Events above PeV, recorded between May 13, 2011 and July 08, 2018, the first seven years after IceCube detector completion. In green, we show the Galactic plane (green dotted line) as well as the $|b|=20\degree$ band 
    (green solid lines). The blue star represents the Galactic Center.  The color of the dots indicates the reconstructed 
    neutrino energy of the neutrino candidate events.}
    \label{fig:events_IC86}
\end{figure}

{\it Methodology and Results.}  We consider all IceCube events (IC86) with track-like signatures and reconstructed 
energy $E \geq \pev$, 
observed during the period May 13, 2011 to July 08, 2018, when the full 
IceCube array was operational~\cite{tmpfix2}. 
All of these 
neutrino candidates are incident on the Earth in the Southern Hemisphere, because for such large energies, neutrino candidates (including atmospheric neutrinos and atmospheric muons) incident 
in the Northern Hemisphere would interact before reaching IceCube.  We only consider neutrino candidates that 
arrive from regions of the sky that are within the HAWC footprint, which encompasses declinations from 
$70^\circ$ to $-30^\circ$~\cite{historical:2023opo}.  In addition, we mask the Galactic plane ($|b|<20\degree$), where a large 
number of gamma ray sources are present.  We are left with 139 sky locations from which neutrino candidates with reconstructed energy
$E \geq \pev$ have been detected by IceCube.  Figure~\ref{fig:events_IC86} shows the  
neutrino candidate events, each considered as a potential source, 
represented by dots, where the color indicates the reconstructed neutrino energy.  The green dotted line shows the location of the Galactic plane, with the masked band ($|b|<20\degree$) outlined in solid green.  The blue star shows the location of the Galactic Center.

\begin{figure}
    \centering
    \includegraphics[width=0.8\textwidth]{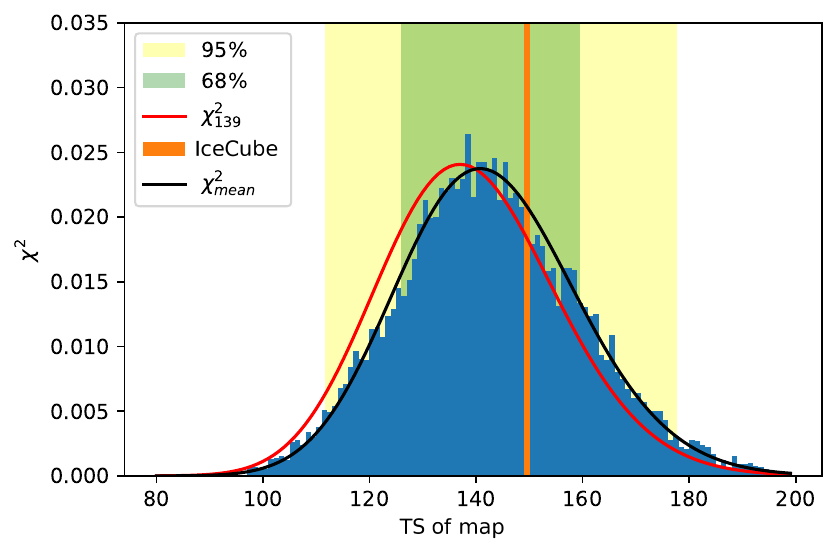}
    \caption{Distribution of the test statistic for $10^4$ random maps (blue) and the result for the 139 sky locations containing PeV neutrino candidates (orange line). A 
$\chi^2$-distribution with 139 degrees of freedom (red curve) and the best-fit $\chi^2$-distribution for 143 degrees of freedom (black curve) are also shown for comparison.  Green and yellow shaded regions represent the 95 and 68\% containment bands for the best fit distribution.} 
    \label{fig:chi_dis}
\end{figure}

We use the publicly-available data from the HAWC 3HWC point source catalog to obtain the signal significance 
for point source detection in each 
pixel, using data from 1523 days of operation~\cite{HAWC:2020hrt}.  The signal significance is the square of a test 
statistic given by $TS = 2 \ln ({\cal L}_{s+b}/{\cal L}_b)$, where ${\cal L}_b$ is the likelihood of the 
HAWC data given the HAWC background model\footnote{{For a point source search, the HAWC collaboration estimates 
background (dominated by misidentified hadronic cosmic rays) using the ``direct integration" method detailed 
in Ref.~\cite{TibetGamma:1999vbk}.  The background rate in any region of sky is found by convolving an isotropic but 
time-dependent background rate with an acceptance factor which depends on the orientation with respect to the 
detector.  Both of these factors are derived from HAWC data.}}, and 
${\cal L}_{s+b}$ is the likelihood of the best fit model of 
background plus a  point source with energy spectrum $\propto E^{-2.5}$, with the normalization as 
a free parameter\footnote{The signal significance is reported with a negative sign if the best fit signal 
model has a negative signal normalization.}.
Since the test statistic in each pixel is $\chi^2$-distributed with one degree of freedom~\cite{HAWC:2020hrt}, the sum of the test statistic 
in 139 pixels should be $\chi^2$-distributed with 139 degrees of freedom.
To demonstrate this, we draw $10^4$ 
sets of 139 random sky locations between declinations $0$ and $-30^\circ$
(excluding the Galactic plane) and histogram the summed test statistics.  In Figure~\ref{fig:chi_dis}, we show the histogram of summed test statistics for the random sky regions (blue) and 
the result for the 139 sky locations containing PeV neutrino candidates (orange line).
For comparison, we also plot a 
$\chi^2$-distribution with 139 degrees of freedom (red curve), and the $\chi^2$-distribution which is a best 
fit to the histogram (black curve).
The best fit $\chi^2$-distribution is for 143 degrees of freedom, similar 
to our expectations for the behavior of the test statistic.\footnote{It is not surprising that there is a slight deviation between the distribution of the summed test statistic and the expected distribution under the background-only hypothesis, since there may be point sources in this region (that is, between declinations of $0$ and $-30^\circ$, excluding the Galactic plane).}  
The green and yellow bands are the $68\%$ and 
$95\%$ containment bands of the best fit $\chi^2$-distribution.  We see that the test statistic obtained from 
the sky locations containing the IceCube neutrino candidates fits well inside the $68\%$ containment band, indicating that 
there is no statistical preference for point sources within these regions.

Note that the HAWC angular resolution varies between $0.1^\circ - 1.0^\circ$, and its pointing uncertainty may be as 
large as $0.3^\circ$ in this region of the sky~\cite{HAWC:2020hrt}.  
{For the IC neutrino sample we use, in the energy range we consider, the median neutrino 
pointing uncertainty is $\sim 0.2^\circ$~\cite{IceCube:2021xar}.  
Our method amounts to a search for excess gamma ray emission from the reconstructed directions of high-energy 
IC neutrino candidate events. This method would not be sensitive to correlations between the neutrino and gamma ray fluxes if 
the reconstructed neutrino direction is sufficiently far from the actual source.  
Since the pointing uncertainties are comparable, a small but non-negligible fraction of $\nu-/\gamma-$sources 
would evade this search in this manner. }

\begin{figure}[h!]
    \centering
    \includegraphics[width=0.8\textwidth]{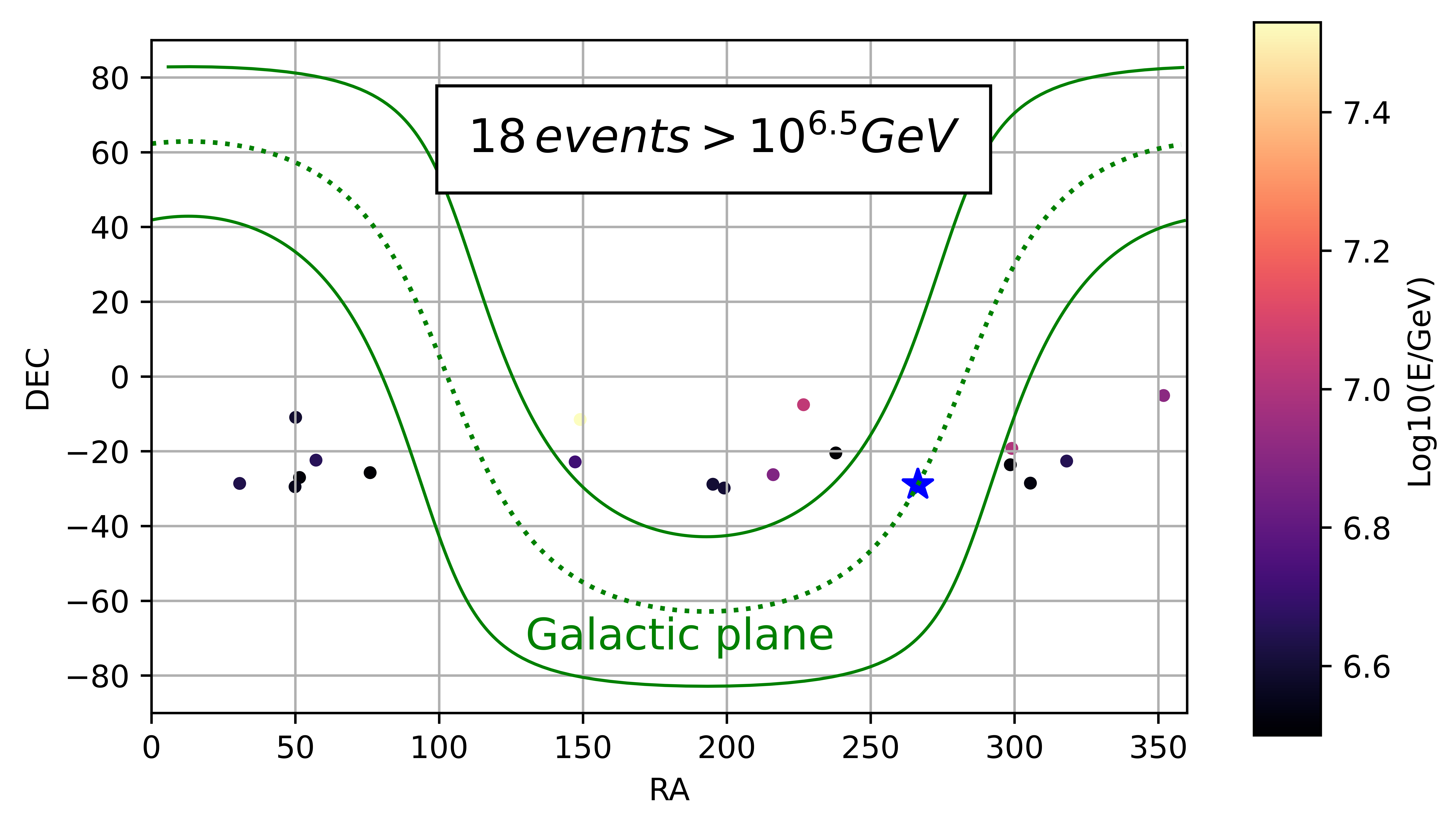}
    \caption{Similar to Figure~\ref{fig:events_IC86}, but for events with 
    $E > 3.2~\pev$.}
    \label{fig:neventsmap_IC86_VII_PeV}
\end{figure}

\begin{figure}[h!]
    \centering
    \includegraphics[width=0.8\textwidth]
{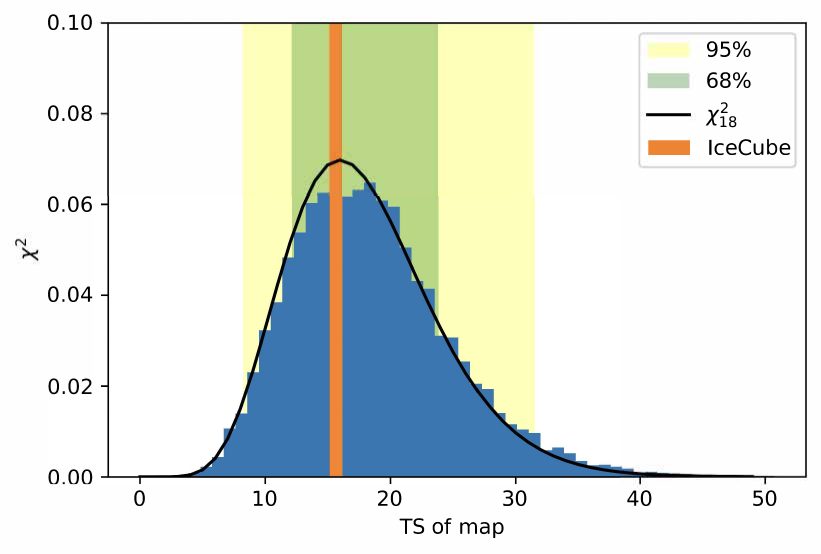}
    \caption{Similar to Figure~\ref{fig:chi_dis}, but restricted to neutrino candidates with 
    reconstructed energy $E > 3.2~\pev$.} 
    \label{fig:aPeV_10K_bands}
\end{figure}

The solid angle encompassed by 139 sky regions of 
size $0.3^\circ$ is small compared to the solid angle of the total sky region from which they are drawn.  
However, if the neutrino threshold were to be lowered significantly, one would obtain so many neutrinos
that the regions of sky (of $0.3^\circ$ radius) covering these neutrino candidates would be an ${\cal O}(1)$ fraction 
of the intersection of the HAWC and IceCube footprints, largely negating the validity of this type of analysis.  
One might ask, however, if there is a correlation between gamma rays seen by HAWC and the highest energy 
neutrino candidates seen by IceCube.  We investigate this possibility 
by performing a similar 
analysis using only the 18 neutrino candidate events detected by IceCube with $E > 3.2~\pev$, which are also within the 
HAWC footprint but outside the mask of the Galactic plane.  
We present the corresponding skymap in Figure~\ref{fig:neventsmap_IC86_VII_PeV}, and the distribution of summed test statistics 
in Figure~\ref{fig:aPeV_10K_bands}.
We again find no significant 
statistical preference for gamma-ray point sources in this combined region.

{\it Conclusion}
We have investigated the possibility of a correlation between high energy neutrino candidate events ($E \geq 1$ PeV)
observed by IceCube and the high energy photons observed by HAWC.  We have used data from the publicly-available 3HWC  
point source catalog and the publicly-available IceCube data to compare
the summed test statistic for gamma ray point sources in the regions of sky  
containing IceCube PeV neutrino candidates to the summed test statistic for the same number of random sky 
regions.  We find that the test statistic for regions of sky containing high energy neutrino candidates lies well within 
the $68\%$ containment region of the background distribution, indicating no statically significant correlation.  

A correlation between high energy neutrinos and high energy photons could arise if there were a population 
of unidentified point sources that produce(d) both neutrinos and photons.  It would be interesting to investigate 
this possibility further with future data and with dedicated analyses.  We have used a publicly-available 
HAWC point source search tool.  However, HAWC's footprint largely covers the Northern Hemisphere, and only a 
fraction of its footprint covers the region of the Southern Hemisphere from which PeV+ neutrinos could be seen 
at IceCube.  As more extensive neutrino and high energy gamma-ray data sets become available, it would be 
useful to apply this technique to data sets with greater overlap.

While this analysis did not find any correlation between gamma-ray sources and high-energy neutrino candidate events, the study presented here has several limitations that a comprehensive analysis, conducted by the experimental collaborations, could overcome, potentially leading to significant sensitivity improvements. 
We do not use neutrino candidate reconstructed energy information other than for the definition of the applied energy thresholds. The reconstructed energy information combined with the declination angle can be used to assign a probability of a neutrino being of astrophysical origin. This information can be included as an event weighting for enhanced sensitivity. Further, the reconstruction uncertainty on the neutrino direction is not taken into account, while a weighting with the likelihood of individual neutrino points of origin would result in further sensitivity improvements.  
The publicly-available HAWC analysis tool assumes that gamma-ray point sources have a power-law energy spectrum with a slope of -2.5.
For a given model of high energy neutrino 
production (for example, from dark matter annihilation), one could also predict the expected gamma ray spectrum.  
Including spectral information in a dedicated analysis could potentially provide greater sensitivity to correlations 
between high energy photons and neutrinos.

{Not all sources of both gamma-rays and neutrinos are amenable to the search we have 
conducted.  In particular, we have searched for spatial correlations between high energy neutrino candidates ($> \pev$) 
seen by IceCube and lower energy photons seen by HAWC.  Depending on the production mechanisms for these messengers, 
the distance to the source, and the environment between the source and Earth, even a source of both gamma rays and neutrinos 
may not produce detectable event rates within the available energy windows for these messengers.  The limited publicly-available 
data thus limits the information one could obtain about the nature of the sources to which a search of this type is sensitive, 
beyond the important fact of their existence.  When more data is publicly available, a future study involving spectral information 
from broader data sets would allow constraints on (or detections of) specific classes of sources and messenger production mechanisms.}

{\bf Acknowledgements.}  
We are grateful to Melissa Diamond for useful discussions.
JK is grateful to the University of Utah for its hospitality.
JK is supported in part by DOE grant DE-SC0010504.
PS is supported in part by NSF grant PHY-2014075.
CR acknowledges support from NSF Grant No. PHY-2309967 and from the National Research Foundation of Korea (NRF) for the Basic Science Research Program NRF-2020R1A2C3008356. 
NT thanks the University of Utah for its support.

\bibliographystyle{unsrt}
\bibliography{ref.bib}

\end{document}